\documentclass[pra,aps,twocolumn,showpacs,epsfig]{revtex4}
\usepackage{color}
\usepackage{graphicx,floatflt,amssymb,epsfig,epsf,amsmath} 

\begin{document}

\title{Stability of attractive bosonic cloud with van der Waals interaction} 

\author{Anindya Biswas$^{1}$, Tapan Kumar Das$^{1}$, 
Luca Salasnich$^{2,3}$ and Barnali Chakrabarti$^{4}$}

\affiliation{$^{1}$Department of Physics, University of Calcutta, 
92 A.P.C. Road, Kolkata 700009, India \\
$^{2}$CNR and CNISM, Unit\`a di Padova, 
Dipartimento di Fisica ``Galileo Galilei'', Universit$\grave{a}$ 
di Padova, Via Marzolo 8, 35122 Padova, Italy \\
$^{3}$CAMTP, University of Maribor, Krevova 2, 2000 Maribor, Slovenia \\
$^{4}$Department of Physics, Lady Brabourne College, P1/2 
Surawardi Avenue, Kolkata 700017, India}

\begin{abstract} 
We investigate the structure and stability of Bose-Einstein condensate 
of $^{7}$Li atoms with realistic van der Waals interaction 
by using the potential harmonic expansion method. Besides the 
known low-density metastable solution with contact delta function interaction, 
we find a stable branch at a higher density which corresponds to the 
formation of an atomic cluster. Comparison with the results of non-local 
effective interaction is also presented. We analyze the effect of trap 
size on the transition between the two branches of solutions. We also 
compute the loss rate of a Bose condensate due to two- and three-body 
collisions.
\end{abstract}



\maketitle

\section{Introduction}

Experiments of Bose-Einstein condensation (BEC) with $^{7}$Li atoms is still a 
challenging research area as the $s$-wave scattering length ($a_{s}$) 
is negative, which indicates an attractive atom-atom interaction [1]. 
A homogeneous condensate with negative scattering length is impossible, 
as it leads to ever increasing density [2]. 
As $a_{s}$ is negative, the attractive interaction 
energy gradually increases  with increase in the number of atoms in a small 
volume and the condensate approaches collapse. However, the situation changes 
drastically in the presence of an external confinement. A spatially confined 
BEC may be stable for a small, finite number of atoms [3]. In the presence of 
a confining potential, the destabilizing effect of the effective negative 
interaction 
is balanced by the kinetic pressure of the gas and a metastable condensate can 
form [1]. For $^{7}$Li, $a_{s}=-(27.3\pm0.8)$ Bohr and for T=0, a metastable 
condensate exists when the number of atoms is less than the critical number, 
which is roughly 1300 [4], whereas theories predict that BEC can occur in a 
trap with no more than about 1400 atoms [3,5-7]. 

During the last decade a number of articles have been published where 
the properties 
and stability of the attractive condensate have been 
discussed in details [3,5-7]. The two- and three-body decay processes near the 
collapse have also been studied in a number of papers [7,8]. However, 
all the earlier 
calculations use the mean-field approximation and the ground state energy is 
calculated by the Gross-Pitaevskii (GP) energy functional~\cite{dalfovo}. 
GP theory is based on the pseudopotential form of the atom-atom interaction 
{\it i.e.} a $\delta$-function potential, where the strength of interatomic 
interaction is absolutely determined by a single parameter $a_{s}$. But due 
to the presence of the pathological singularity at $|{\vec x}|=0$, 
the Hamiltonian becomes unbound from below and it has been 
emphasized previously that a 
$\delta$-function is not suitable as an exact potential in 3D attractive 
systems~\cite{geltman,gao}. 
Again, as the attractive BEC becomes highly correlated near the 
critical point, uncorrelated GP equation can not take care of 
the effect of interatomic correlation. 

Thus the  motivation of the present work is to study the same system using an 
{\it{ab initio}} many-body method, that is 
the potential harmonic expansion method \cite{ballot,fabre}, 
with the incorporation of a realistic potential, 
like the van der Waals potential. The presence of a hard sphere below some 
cutoff 
radius and the $-C_{6}/|{\vec x}|^{6}$ tail properly 
take care of the effects of 
both short-range and long-range correlation and accurately represent the 
realistic 
features. We find the commonly known low-density metastable branch and discuss 
its stability, structure and decay rates due to inelastic two- and three-body 
collisions. Besides the known solution, we also find another branch of 
solution 
at a higher density. Due to the use of the realistic van der Waals interaction 
in the many-body calculation, we have a deep attractive well on the left side 
of the metastable region in  the effective potential. In the standard GP theory, 
the metastable region just vanishes at the critical point and the whole 
condensate 
collapses into the singular well and the fate of the condensate is not 
predicted further. However, in our calculation the metastable condensate leaks 
through the intermediate barrier and settles down in the extremely narrow 
(width $\sim 0.05\mu m$) well. The atoms become highly correlated and due to 
high two-body and three-body collisions, atoms form a cluster. 
This corresponds to the high-density branch. In the present communication, 
we investigate the transition between the two branches of solutions as a 
function of the number of atoms and their dependence on the trap size. 
We also calculate decay rates of the condensate due to two- and three-body 
collisions. In other articles~\cite{luca}, the existence of a similar type 
of new stable branch has been reported, which is intermediate in density 
between 
the dilute metastable state and the collapsed state. This is described as the 
effect of use of non-local interaction in the GP theory. It has been pointed 
out that for $^{7}$Li system, the scattering cross-section has a momentum dependence 
and the effective interaction is non-local, changing from attractive to 
repulsive at a characteristic range $r_{e}$. They calculated the 
properties of the attractive condensate by the variational technique. 
Using a Gaussian trial wavefunction they minimized the quantum action. 

In the present communication, we also compare our results with those 
obtained using the non-local interaction. A road map of the present 
study is given below. Sec.II contains the many-body approach used in 
this work, which is based on the potential harmonic 
expansion method. Sec.III 
presents the calculation with a non-local interaction. Sec. IV contains a  
comparison of the structure and stability of the bosonic system of both the low- and 
high-density branches obtained from two different theoretical calculations 
and the discussion of the many-body effect. Sec.V provides a summary of our 
results and their conclusions.\\

\section{Many-body calculation: (a) Potential harmonics basis}

We consider a system of A=(N+1) identical spinless bosons (each of mass m) 
confined in an external trapping potential $U_{trap}({\vec x}_j)$ 
and interacting via 
a two-body potential V($\vec{x}_{i}-\vec{x}_{j}$), $\vec{x}_i$ being the 
position vector of the $i$-th particle. The corresponding Schr\"{o}dinger 
equation is written as 
\begin{eqnarray} 
\Big[ &-&\frac{\hbar^2}{2m} \sum_{i=1}^{A} \nabla_{i}^{2} + 
\sum_{i=1}^{A}U_{trap}(\vec{x}_{i}) +
\nonumber 
\\
&&\sum_{i,j>i}^{A} V(\vec{x}_{i}
-\vec{x}_{j}) - E \Big] \psi(\vec{x}_{1},...,\vec{x}_{A})=0.
\end{eqnarray}
We define N Jacobi vectors as 
\begin{equation}
 \vec{\zeta}_{i}=\sqrt{\frac{2i}{i+1}}\left( \vec{x}_{i+1} - \frac{1}{i} 
\sum_{j=1}^{i} \vec{x}_{j}\right) , \hspace*{0.5cm} (i=1,...,N) . 
\end{equation}
The center of mass is $\vec{R}=\sum_{i=1}^{A}\vec{x}_i/A$. 
Since the labelling of particles is arbitrary, we choose the relative 
separation of $(i,j)$ interacting pair ($\vec{x}_{ij}=\vec{x}_i - 
\vec{x}_j$) as $\vec{\zeta}_{N}$. 
We next define the hyperradius $r$ of the set of N Jacobi vectors
through~\cite{ballot}
\begin{equation}
r^{2}= \sum_{i=1}^{N} \zeta_{i}^{2} = \frac{2}{A} 
\sum_{i,j>i} x_{ij}^{2} =2\sum_{i=1}^{A}r_{i}^{2} \hspace*{0.3cm},
\end{equation}
where $\vec{r}_{i}=\vec{x}_i-\vec{R}$ is the position vector of the 
$i$-th particle 
from the center of mass of the system. 
In this way the relative motion of the bosons is described by
\begin{eqnarray} 
\Big[&-&\frac{\hbar^{2}}{m} \sum_{i=1}^{N} \nabla_{\zeta_{i}}^{2}+
V_{trap}(r) + 
\nonumber
\\
&&V_{int}(\vec{\zeta}_{1}, ..., \vec{\zeta}_{N})-E_{R} \Big] 
\psi(\vec{\zeta}_{1}, ..., \vec{\zeta}_{N}) = 0 ,
\end{eqnarray} 
where E$_{R}$ is the energy of the relative motion.   
In the present work 
$U_{trap}({\vec x}_i) = \frac{1}{2}m\omega^{2}|{\vec x}_i|^{2}$ 
is a spherically symmetric harmonic oscillator potential and consequently 
\begin{equation}
\sum_{i=1}^{A} U_{trap}({\vec x}_i) 
=\frac{1}{4} m\omega^{2} r^{2}+\frac{1}{2}mA\omega^2R^2.
\end{equation} 
The first term on the right side is the effective trap potential 
$V_{trap}(r)$ for 
the relative motion and the second term is the trap potential of the 
center of mass motion. The equation for the center of mass motion 
separates completely and is simply the equation for a three dimensional 
harmonic oscillator. The total energy of the system is thus 
$(E_R+\frac{3}{2}\hbar \omega)$. 
In Eq.~(4), $V_{int}$ is the sum of all 
pair-wise interactions, expressed in terms of the Jacobi vectors.
In our approach we consider that when $(ij)$ pair 
interacts, the rest of the bosons are inert spectators. 
So we define a 
hyperradius $\rho_{ij}$ for the remaining (N-1) noninteracting 
bosons as~\cite{fabre}
\begin{equation}
 \rho_{ij} = \left[ \sum_{k=1}^{N-1}\zeta_{k}^{2}\right] ^{1/2},
\end{equation}
so that $r^{2}=x_{ij}^{2}+\rho_{ij}^{2}$. The hyperangle $\phi$ is 
introduced through $x_{ij}=r \cos\phi$ and $\rho_{ij}=r \sin\phi$. 
Besides $r,\phi,\vartheta,\varphi$ (where $\vartheta$ and $\varphi$ 
are the polar angles of $\vec{x}_{ij}$), there are $(3N-4)$ remaining 
variables. These are constituted by $2(N-1)$ polar angles associated with 
($\vec{\zeta}_{1},...,\vec{\zeta}_{N-1}$) and $(N-2)$ angles defining 
their relative lengths and collectively denoted by $\Omega_{N-1}^{(ij)}$,  
called hyperangles in the $3(N-1)$-dimensional space. The corresponding 
form of Laplace operator can be found in~\cite{ballot}. In the potential 
harmonics expansion method (PHEM), $\psi$ is decomposed 
into Faddeev components, 
$\phi_{ij}$ for the $(ij)$ interacting pair as,
\begin{equation}
 \psi=\sum_{i,j>i}^{A}\phi_{ij}.
\end{equation}
Then Eq.~(4) can be written as 
\begin{equation}
 (T+V_{trap}-E_{R})\phi_{ij}=-V(x_{ij})\sum_{k,l>k}^{A} \phi_{kl},
\end{equation}
where $T=-\frac{\hbar^{2}}{m}\sum_{i=1}^{N} \nabla^{2}_{\vec{\zeta}_{i}}$. 
Note that the assumptions that correlations higher than two-body ones in 
$\psi$ are negligible and that the angular and hyperangular momenta of the 
system are contributed by the interacting pair only~\cite{chak1}, 
make the Faddeev 
component $\phi_{ij}$ independent of the coordinates of all the particles, 
other than the interacting pair~\cite{ballot}. Thus $\phi_{ij}=\phi_{ij}
(\vec{x}_{ij},r)$. With this assumption we expand $\phi_{ij}$ in the subset 
of hyperspherical harmonics (HH) necessary for the expansion of 
V($\vec{x}_{ij}$). Thus,
\begin{equation}
 \phi_{ij}(\vec{x}_{ij},r)=r^\frac{-(3N-1)}{2}
\sum_{K}\mathcal{P}_{2K+l}^{lm}(\Omega^{ij}_{N})u_{K}^{l}(r).
\end{equation}
This subset of HH is called the potential harmonics (PH) basis and is 
denoted by $\{\mathcal{P}_{2K+l}^{lm}(\Omega^{ij}_{N})\}$; $\Omega^{ij}_{N}$ 
denotes the set of hyperangles in 3N dimensional space with the interacting 
pair uniquely identified as above. Note also that $\mathcal{P}_{2K+l}^{lm}
(\Omega^{ij}_{N})$ is independent of ($\vec{\zeta}_{1},...,\vec{\zeta}_{N-1}$) 
and thus no contribution to the orbital and hyper angular momenta of the 
condensate comes from the remaining $(N-1)$ noninteracting bosons. Thus 
the total orbital angular momenta of the condensate and its projection 
are simply those of the interacting pair, {\it viz.} $l$ and $m$ respectively. 
The complete analytic expression for $\mathcal{P}_{2K+l}^{lm}
(\Omega^{ij}_{N})$ can be found in~\cite{fabre}. Substitution of Eq.~(9) 
into Eq.~(8) and subsequent projection on a particular PH gives~\cite{chak1} 
\begin{eqnarray}
\Big[&-&\frac{\hbar^{2}}{m}\frac{d^{2}}{dr^{2}}+\frac{\hbar^{2}}{m}
\frac{{\cal L}_{K}({\cal L}_{K}+1)}{r^{2}} + V_{trap}(r) -E_{R} \Big] 
u_{K}^{l}(r) 
\nonumber 
\\
&+& \sum_{K^{\prime}} f_{K^{\prime}l}^{2}V_{KK^\prime}(r)
u_{K^{\prime}}^{l}(r)=0 , 
\end{eqnarray}
where $V_{KK'}(r)$ is the potential matrix and is given by,
\begin{equation}
 V_{KK^{\prime}}(r) = \int {\mathcal P}_{2K+l}^{{lm}^{*}}(\Omega^{ij}_{N})
V(x_{ij})
{\mathcal P}_{2K^{\prime}+l}^{lm}(\Omega^{ij}_{N})d\Omega_{N}^{ij}.
\end{equation}
The quantities $\mathcal{L}_{K}$ and $f^{2}_{Kl}$ are given by
\begin{equation}
 \begin{array}{rcl}
{\cal L}_{K} &=& 2K+l+\frac{3N-3}{2}\\
f_{Kl}^{2} &=& \sum_{k,l>k} <{\mathcal P}_{2K+l}^{lm}(\Omega^{ij}_{N})
|{\mathcal P}_{2K+l}^{lm}
(\Omega^{kl}_{N})>,\\
\end{array}
\end{equation}
the latter being the overlap of the PH for the $(ij)$-partition 
(corresponding to only the $(ij)$-pair interacting) with the 
sum of PH for all partitions. 
The quantum number $K$ in Eq.~(10) is the hyperspherical 
(grand orbital) quantum number in the $3N$ dimensional 
space~\cite{ballot,fabre}, 
analogous with the orbital angular momentum in three dimensional 
space. The ordinary orbital angular momentum of the system 
and its projection are $l$ and $m$, as indicated earlier. All 
other intermediate angular momentum quantum numbers of the system 
take zero eigen values, due to the choice of the PH basis. 
Thus the choice of the PH basis reduces the complications 
of the many-body system immensely.
Eq.~(10) can be put in a symmetric form as 
\begin{eqnarray}
\Big[&-&\dfrac{\hbar^{2}}{m} \dfrac{d^{2}}{dr^{2}} + \dfrac{\hbar^{2}}{mr^{2}}
\{ {\cal L}({\cal L}+1) + 4K(K+\alpha+\beta+1)\} 
\nonumber 
\\
&+& V_{trap}(r) -E_R \Big] U_{Kl}(r) 
\\
&+& \sum_{K^{\prime}}f_{Kl}V_{KK^{\prime}}(r)f_{K'l} U_{K^{\prime}l}(r) = 0 ,
\nonumber
\end{eqnarray}
where ${\cal L}=l+(3A-6)/2$, 
$U_{Kl}(r)=f_{Kl}u_{K}^{l}(r)$, $\alpha=\frac{3A-8}{2}$ and 
$\beta=l+\frac{1}{2}$. 

\section{Many-body calculation: 
(b) Incorporation of two-body correlation function}

In the experimentally achievable BEC, the average interparticle separation 
is much larger than the range of two-body interaction. This is indeed 
required to prevent atomic three-body collisions and formation of 
molecules. Moreover, the energy of the interacting pair is extremely 
small. Thus the two-body interaction is generally represented by the 
$s$-wave scattering length ($a_{s}$). 
In the GP picture, the effective interaction is given in terms of 
$a_s$ alone, making the results independent of the shape of the 
interatomic potential.
Hence, a positive value of $a_{s}$ gives a repulsive condensate and a 
negative $a_{s}$ gives an attractive condensate. 
However, a realistic interatomic interaction, like the van der Waals 
potential, is always associated with an attractive 
$-1/|{\vec x}_{ij}|^6$ tail at larger separations 
and a strong repulsion at short separations. 
Depending on the nature of 
these two parts, $a_s$ can be either positive or negative~\cite{pethick}. 
For a given two-body potential $V(x_{ij})$ having a finite range, 
$a_s$ can be obtained 
by solving the zero-energy 
two-body Schr\"odinger equation for the wave function $\eta(x_{ij})$
\begin{equation}
-\frac{\hbar^2}{m}\frac{1}{x_{ij}^2}\frac{d}{dx_{ij}}\left(x_{ij}^2
\frac{d\eta(x_{ij})}{dx_{ij}}\right)+V(x_{ij})\eta(x_{ij})=0 . 
\end{equation}
The asymptotic part of $\eta(x_{ij})$ has the form $(c_1x_{ij}+c_2)$ and the 
corresponding $s$-wave scattering length is given by $a_s=-\frac{c_2}{c_1}$.

Now the rate of convergence of the PH expansion in Eq.~(9) is 
very slow. This can be understood from the fact that the leading term of this 
basis, {\it viz.} the term with $K=0$, is a non-zero constant, while 
$\phi_{ij}$ 
must be vanishingly small for small values of $x_{ij}$ due to the strong 
short-range repulsion of $V(x_{ij})$. Consequently a large number of terms 
is needed to represent $\phi_{ij}$ faithfully for small values of $x_{ij}$.
The rate of convergence is improved dramatically by introducing a short-range 
correlation function in the expansion basis~\cite{CDLin}, that represents 
the true nature of $\phi_{ij}(x_{ij})$ as $x_{ij} \rightarrow 0$. In the 
experimental BEC, the energy of the interacting pair 
(which is $\sim \hbar \omega$) is negligible compared 
with the depth of interaction potential $V(x_{ij})$ (which is of the 
order of typical atomic energy scale). Thus $\eta(x_{ij})$ 
is a good representation of the short-range behavior of $\phi_{ij}(x_{ij})$. 
Hence we use $\eta(x_{ij})$ as a short-range correlation function in the 
PH expansion, to enhance its rate of convergence~\cite{chak2}: 

\begin{equation}
\phi_{ij}(\vec{x}_{ij},r)=r^\frac{-(3N-1)}{2}\sum_{K}\mathcal{P}_{2K+l}^{lm}
(\Omega^{ij}_{N})u_{K}^{l}(r)\eta(x_{ij}).
\end{equation}
As $\eta(x_{ij})$ correctly reproduces the short separation behavior  
of the interacting-pair Faddeev component, convergence rate of 
the PH expansion for the 
residual part of $\phi_{ij}$ is quite fast. Validity of this 
statement has been tested in numerical calculations. In an actual 
calculation, we solve Eq.~(14) for the zero energy two-body wave function 
$\eta(x_{ij})$ in the chosen two-body potential $V(x_{ij})$, after 
adjusting the hard core radius ($r_c$), such that $a_{s}$ has the 
desired value~\cite{Das1}. This $\eta(x_{ij})$ is then used in Eq.~(15).

Then the potential matrix becomes  
\begin{eqnarray}
V_{KK^{\prime}}(r) &=& (h_{K}^{\alpha\beta}
h_{K^{\prime}}^{\alpha\beta})^{-\frac{1}{2}}
\int_{-1}^{+1} \Big\{ 
P_{K}^{\alpha 
\beta}(z)
V\left(r\sqrt{\frac{1+z}{2}}\right) 
\nonumber 
\\
&&P_{K^{\prime}}^{\alpha \beta}(z)\eta\left(r\sqrt{\frac{1+z}{2}}\right)
w_{l}(z) \Big\} dz .
\end{eqnarray}
where $P_{K}^{\alpha\beta}(z)$ is the Jacobi polynomial, whose norm and 
weight function are $h_{K}^{\alpha\beta}$ and $w_{l}(z)$ 
respectively~\cite{abram}. We restrict the K-sum in Eq.~(15) to an upper 
limit $K_{max}$ from the requirement of convergence. Finally we solve the 
set of coupled differential equations (CDE), Eq.~(13), by hyperspherical 
adiabatic approximation~\cite{das2}. The total energy of the system 
is obtained by adding energy of the center of mass motion 
($3\hbar \omega/2$) to $E_R$. The wave function of the condensate in 
the physical space can be obtained from $U_{Kl}(r)$, using Eqs.~(7) and 
(9) and the ralations between hyperspherical and physical coordinates.  

\section{GP theory with non-local interaction}

In the standard treatment of alkali-metal atoms, the interatomic interaction 
is chosen as a local form, i.e., momentum independent. The effective zero-range
 pseudopotential has the form $V({\vec x})=g\delta^{3}({\vec x})$, where 
$g=\frac{4\pi\hbar^{2}a_{s}}{m}$. 
However, this is not correct when the scattering cross-section has a 
significant momentum dependence. This implies that the effective interaction 
is non-local, changing from attractive to repulsive at a characteristic 
range $r_{e}$. The system of $^{7}$Li atoms exhibit such momentum dependence 
and the effective $^7$Li-$^7$Li interaction can be written as [11] 
\begin{equation}
V(k)=\frac{4\pi\hbar^{2}}{m}[a_{r}+(a_{s}-a_{r})f(kr_{e})]
\end{equation}
where $a_{s}$ corresponds to the attractive potential and has the value 
$a_{s}=-27.3$ Bohr, whereas the repulsive contribution is modeled by a 
local positive term with $a_{r}=6.6$ Bohr and $r_{e}$ = $10^{3}$ Bohr. 
The shape function $f(x)$ may be chosen as a Lorenzian 
$f(x)=(1+x^{2})^{-1}$. In real space the effective inter-atomic 
potential then reads 
\begin{equation}
V({\vec x}) = \frac{4\pi\hbar^{2}a_s}{m} \delta^{3}({\vec x}) + 
\frac{\hbar^{2}(a_s-a_r)}{m r_e^2} \ {e^{-|{\vec x}|/r_e} 
\over |{\vec x}|} \; . 
\end{equation}
This potential appears in the nonlocal GP energy functional 
\begin{eqnarray}
E = \int \phi^*(\vec{x}) 
\left[ -{\hbar^2\over 2m} \nabla^2 + U_{trap}(|{\vec x}|)
\right] \phi(\vec{x}) \ d^3\vec{x} 
\nonumber 
\\
+ {1\over 2} \int |\phi({\vec x})|^2 
V({\vec x}-{\vec x}') |\phi({\vec x}')|^2  \ d^3\vec{x} \ d^3\vec{x}' \; , 
\end{eqnarray}
where $\phi({\vec x})$ is the spherically-symmetric GP wave function. 
In earlier calculations~\cite{luca} 
the variational ansatz of the wave function is chosen as 
\begin{equation}
 \phi(|\vec{x}|)=N^{1/2}\frac{1}{\pi^{3/4}\sigma^{3/2}}{\rm exp}
\left\lbrace -\frac{|{\vec x}|^{2}}{2\sigma^{2}a_0^2}\right\rbrace 
\end{equation}
where $\sigma$ is the variational parameter, 
$\sigma a_0$ being the width of the condensate 
with $a_{0}= \sqrt{\frac{\hbar}{m\omega}}$ the characteristic length of 
the harmonic trap. In this way the energy $E$ of the system becomes~\cite{luca}
\begin{equation}
E(\sigma ) = {N \hbar \omega\over 4} \left[  
\frac{3}{\sigma^{2}}+3\sigma^{2}+2N \left( \frac{\gamma_{r}}
{\sigma^{3}}+\frac{\tau_{1}}{\sigma}-\tau_{2}g(\chi\sigma) \right) 
\right] \; , 
\end{equation}
with $\gamma_{r}=(2/\pi)^{1/2}a_{r}/a_{0}$, $\tau_{1}=(2/\pi)^{1/2}a_{0}
(a_{s}-a_{r})/r_{e}^{2}$, 
$\chi=2^{-1/2}a_{0}/r_{e}$, $\tau_{2}=a_{0}^{2}(a_{s}-a_{r})/r_{e}^{3}$ and 
$g(x)={\rm exp}(x^2)[1-{\rm erf}(x)]$, ${\rm erf}(x)$ is the error function.  
The extrema of $E(\sigma)$ 
are obtained from the following expression,
\begin{equation}
 N=(1-\sigma^{4}){\left[ -\gamma_{r}\sigma^{-1} - \frac{\tau_{1}\sigma}{3} + 
\frac{2\chi\tau_{2}\sigma^{3}}{3\sqrt{\pi}} - \frac{2\chi^{2}\tau_{2}\sigma^{4}
g(\chi\sigma)}{3} \right]}^{-1} 
\end{equation}
which actually gives the number of bosons as a function of the scaled 
size $\sigma$ of the cloud. Depending on the choice of parameters 
and the number of atoms, this equation either has one root 
or three roots. When three solutions are present, 
one corresponds to the low-density metastable solution (which we expect for 
local interaction), one corresponds to the high density 
stable solution (which is the 
effect of non-locality) and the middle one corresponds to an unstable state 
({\it i.e.} local maximum). The variational results for different 
trap sizes are discussed in the result section.

\section{Results}

We consider the experiment on $^{7}$Li condensate at Rice 
University~\cite{iju}, for which $a_{s}=-27.3$ Bohr. 
We take $a_{0}$ as the unit of
length and energy is expressed in units of the oscillator energy quanta 
$\hbar\omega$. The trap size 
corresponding to the Rice University experiment is $a_{0}=3.0 \mu m.$ 
For the many-body calculation, 
we choose the van der Waals potential, whose short range repulsion is 
characterized by a hard core of 
radius $r_{c}$ and the long range part is described by an attractive tail 
$-C_6/|{\vec x}_{ij}|^6$, where 
$C_{6}$ is the strength parameter which is known from experiments. 
Choice of a realistic interatomic interaction like the van der Waals 
potential is very important, as it has been shown that the 
shape-independent approximation is not valid in typical laboratory 
BECs~\cite{chak3}. 
In the local mean-field description the 
two-body interaction is usually represented by the $s$-wave scattering 
length $a_{s}$. To mimic the 
$^{7}$Li trap of Rice University, our chosen parameters are 
$C_{6}=1.71487\times10^{-12}$~o.u. and 
$r_{c}=5.3378\times10^{-4}$~o.u., which reproduce the experimental 
value of $a_s$. With these sets of 
parameters we solve the set of coupled differential equations by the 
hyperspherical adiabatic 
approximation (HAA)~\cite{das2}. We assume that the hyperradial motion is 
slow in comparison with the hyperangular motion. 
Hence, for a fixed value of $r$, the equation for 
the hyperangular motion can be solved adiabatically. The energy eigen 
value of this equation is a parametric function of $r$ and provides an 
effective potential for the hyperradial motion. In the HAA prescription, 
the lowest lying such potential is used for the ground state of the 
system. Although Eq.~(13) involves the hyperradius $r$ only, the 
hyperangular motion appears through the coupling matrix 
$V_{Kk^{\prime}}(r)$. Solution of the hyperangular motion for a 
fixed value of $r$ is equivalent to diagonalizing the hyperangular 
Hamiltonian in the potential harmonics basis, {\it i.e.} diagonalizing 
the potential matrix $V_{KK^{\prime}}(r)$ together with the 
hypercentrifugal term of Eq.~(13). 
The lowest eigenvalue of this matrix, $\omega_{0}(r)$, is the 
effective potential in the hyperradial space, in which collective motion 
of the condensate takes place. 
Thus the energy and wave function of the condensate is obtained by 
solving the equation for hyperradial motion 
\begin{equation}
 \left[ -\frac{\hbar^{2}}{m}\frac{d^{2}}{dr^{2}} + \omega_{0}(r)
+ \sum_{K=0}^{K_{max}} |\frac{d\chi_{K0}(r)}{dr}|^{2}-E_R \right] \zeta_{0}(r)=0
\hspace*{0.3cm},
\end{equation}
where the third term is the overbinding correction term~\cite{das2} and 
$\chi_{K0}$ is the K-th component of the column vector corresponding 
to the eigenvalue 
$\omega_{0}(r)$. Energy and wave function of the system are obtained 
by solving Eq.~(23) numerically, subject to appropriate boundary 
conditions. The partial wave $U_{Kl}(r)$ appearing in Eq.~(13) is 
given in HAA as $U_{Kl}(r) \approx \zeta_0(r)\chi_{K0}(r)$~\cite{das2}.

\vskip 0.5cm 
\begin{figure}[hbpt]
\vspace{-10pt}
\centerline{
\hspace{-3.3mm}
\rotatebox{0}{\epsfxsize=9cm\epsfbox{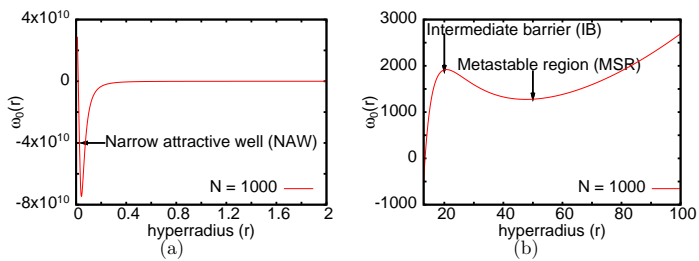}}}
\caption{(Color online) Plot of effective potential $\omega_0(r)$ in 
o.u. against $r$ in o.u. 
for 1000 $^{7}$Li atoms with $a_{0}=3.0 \mu m$. 
Panel (a) shows the narrow attractive well, while panel 
(b) displays the metastable region in detail. Note the different 
horizontal and vertical scales in the two panels.}
\end{figure}

In Fig.~1 we plot the effective potential, $\omega_{0}(r)$ for N~=~1000 
atoms of $^{7}$Li, the 
condensate being metastable as N$<$N$_{cr}$. For N$<$N$_{cr}$, 
the intermediate metastable region (MSR) of $\omega_{0}(r)$ is bounded by 
the high wall of the external trap on the right side and on the left by an 
intermediate barrier (IB) [panel (b)]. On the left of the IB, a very deep 
narrow attractive well (NAW) appears which is again followed by a steeply 
increasing repulsive wall, as $r$ decreases [panel (a)]. The existence of 
a strong repulsive core near the origin ($r\rightarrow0$) is in sharp 
contrast to the GP mean-field picture with local interaction and is the 
immediate reflection of the use of realistic van der Waals 
interaction in quantum many-body calculation. 
In the GP picture, negative 
$a_{s}$ presents a singular attraction (with no repulsive core) near the 
origin. The combination of repulsive core and NAW in the present approach, 
prevents the Hamiltonian from being unbound from below. As the number of 
atoms increases towards the critical number, the height of the intermediate 
barrier decreases, the NAW becomes more negative and narrow.
At N = N$_{cr}$, 
the MSR along with the intermediate barrier disappear, with the local maximum  
of the IB and the local minimum of the MSR  merging to form a point of 
inflexion~\cite{Das1}. 
The value of N where this occurs is N$_{cr}$.
As N becomes larger 
than N$_{cr}$, all the atoms get trapped in the NAW, giving rise to a 
cluster state resulting from the increased interatomic correlations and 
two- and three-body collisions. The release of binding energy is observed 
as a burst of energy in experiments. This is the collapse scenario in the 
present approach. 

Thus the qualitative features of 
the pre-collapse scenario are in fair agreement with the GP local mean-field 
picture away from the critical number. When N = N$_{cr}$, the energy per 
particle becomes $-\infty$ in the local GP method, and the Hamiltonian becomes 
unbound from below. However, in our many-body calculation the Hamiltonian 
has a lower bound due to the presence of a hypercentrifugal barrier as 
$r\rightarrow0$ along with the use of a realistic interatomic interaction 
with a repulsive core of finite size. In the post-collapse state the atoms 
accumulate in the narrow attractive well and form clusters.

\vskip 0.5cm
\begin{figure}[hbpt]
\vspace{-10pt}
\centerline{
\hspace{-3.3mm}
\rotatebox{0}{\epsfxsize=9cm\epsfbox{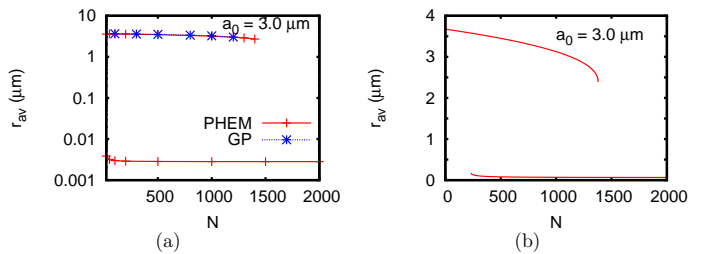}}}
\caption{(Color online) Size of the condensate as a function 
of the number of $^{7}$Li 
atoms using PHEM with van der Waals interaction and GP with local 
interaction (having same $a_{s}$) [panel (a)],
 and GP with non-local interaction [panel (b)].}
\end{figure}

We calculate the size of the condensate as a function of the number of 
$^{7}$Li atoms, for the choice of the parameters mentioned earlier. 
We define the average size of the condensate ($r_{av}$) as the root mean 
square distance of individual atoms from the center of mass and is 
given by~\cite{Das1}
\begin{equation}
r_{av}=\left\langle \frac{1}{A}\sum_{i=1}^{A}(\vec{x}_{i}-\vec{R})^{2} 
\right\rangle ^{1/2}=\frac{<r^{2}>^{1/2}}{\sqrt{2A}}
\end{equation}
where $\vec{R}$ is the center of mass coordinate. We present average size 
of the condensate calculated by PHEM 
in panel (a) of Fig.~2. The upper branch corresponds to the metastable 
condensate. With increase in 
particle number, the total attractive interaction increases as 
$\frac{N(N-1)}{2}$, the system contracts 
and $r_{av}$ decreases as expected. 
For comparison we present in panel (a) the results of the GP equation 
with contact interaction, i.e. with local interaction. 
The lower branch, which corresponds to the  high-density stable branch 
in the deep attractive well,  starts from N=20. In the GP mean-field 
approach with local interaction, there is no other stable branch after 
collapse as the whole condensate falls into the singular well. In our 
calculation we find that the size of the condensate in the deep attractive 
well is of the order of 0.003 $\mu m$, which is the order of the 
size of the atomic 
cluster. This clearly indicates that the metastable condensate forms 
clusters due to high two-body and three-body collisions within an 
extremely narrow well of width $\sim$ 0.05 $\mu m$. However, the 
transition from upper branch to lower branch is discontinuous. 
In panel (b) of Fig.~2, we present results for the same trap but 
with non-local interaction as described in Sec.III. The quantity  
$\sigma a_0$ in Eq.~(20) is basically the average size of the 
condensate and is denoted by $r_{av}$ in panel (b) of Fig.~2. 
The variational parameter $\sigma$ is calculated as a 
function of the number of bosons 
using the algebraic equation~[Eq.~(22)]. The effective potential energy 
$E(\sigma)$ [Eq.~(21)] has  the same qualitative features as the
 many-body effective potential $\omega_{0}(r)$. Unlike the GP effective 
potential with local interaction, (having one local minimum), the use of 
non-local interaction in the calculation of effective potential offers an 
additional absolute minimum together with the local minimum. This absolute 
minimum corresponds to the high-density stable solution whereas the local 
minimum corresponds to the usual metastable solution. The transition 
between the low-density and high-density branches is again discontinuous 
as observed in many-body calculation. However, unlike panel (a) of Fig.~2, 
where the high-density stable branch appears from quite small values of 
N~=~20, in panel (b), the lower branch starts from N~$\simeq$~234. 
This disagreement is due to the fact that the origin of non-locality and 
the occurrence of absolute minimum is different in the two approaches. 
In the many-body calculation the effect of non-locality is inherent due 
to the presence of hypercentrifugal repulsion (arising from the many-body 
treatment) and the use of van der Waals 
potential with a finite range hard core part whereas in the other approach 
the non-locality is created by assuming that the attractive potential has 
a finite range $r_{e}$ together with a local positive term defined by 
high-energy scattering length $a_{r}>0$. This is reflected by the presence 
of a non-local minimum in the effective interaction.

\vskip 0.5cm 
\begin{figure}[hbpt]
\vspace{-10pt}
\centerline{
\hspace{-3.3mm}
\rotatebox{0}{\epsfxsize=9cm\epsfbox{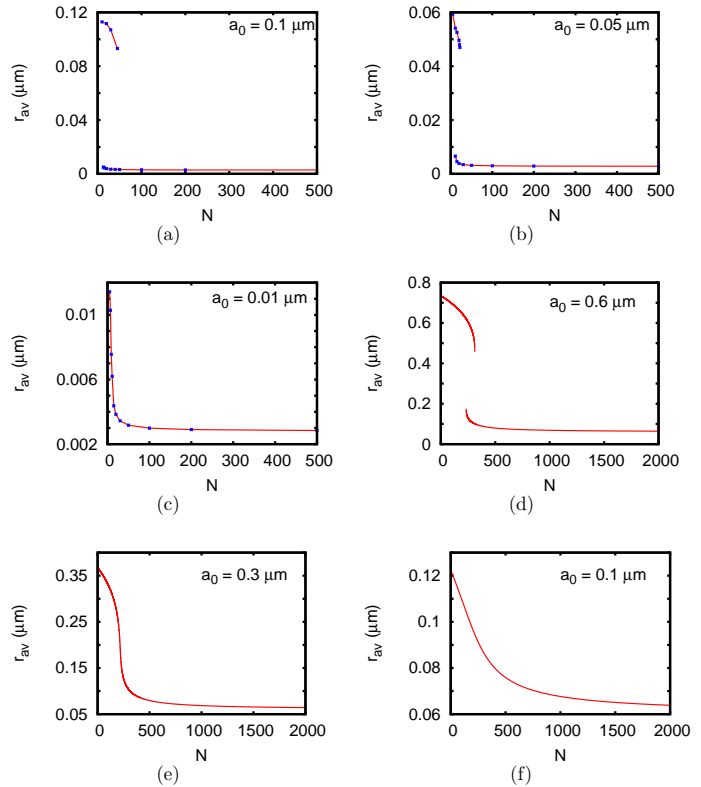}}}
\caption{(Color online) Size of the condensate as a function 
of the number of $^{7}$Li 
atoms using PHEM [panels (a) -- (c)] and GP with non-local interaction 
[panels (d) -- (f)]. Note that in panels (a) - (c), the horizontal 
axis is upto $N=500$, to show the small $N$ part prominently. Calculated 
values of $r_{av}$ for $N$ above $500$ remain practically constant for PHEM.}
\end{figure}

Next, to study the nature of discontinuity between the two branches of 
solution, we repeat our calculation for other trap sizes. We plot the PHEM 
results in panels (a) -- (c ) of Fig.~3 with $a_{0}=0.1 \mu m$ 
($\omega=907.88$kHz), $a_{0}=0.05 \mu m$ ($\omega=3.63$MHz) 
and $a_{0}=0.01 \mu m$ ($\omega=90.79$MHz). 
We observe that by reducing the trap size the discontinuity is strongly 
reduced and at $a_{0}=0.01 \mu m$, the discontinuity disappears. There is 
a smooth evolution from a dilute cloud to a less dilute cloud. We also observe 
that the size is almost independent of the trap size for $N>200$, as 
expected for clusters. For comparison, in panels (d) -- (f) of Fig.~3, 
we plot the variational results with non-local interaction for various 
trap sizes $a_{0}=0.6 \mu m$, $a_{0}=0.3 \mu m$ and $a_{0}=0.1 \mu m$. \\
We get almost same qualitative pictures as shown in panel (a) -- (c) of 
Fig.~3; reducing the trap size, the discontinuity is reduced and at 
$a_{0}=0.1 \mu m$, the metastable branch disappears. However, the 
quantitative disagreement between many-body results and variational 
results with non-local interaction is again attributed to the different 
origin of non-locality in the effective potential. In the variational 
calculation with non-local interaction, the discontinuity disappears at 
$a_{0}=0.1 \mu m$ whereas the corresponding trap size in the many-body 
calculation is $a_{0}=0.01 \mu m$. However, in both cases we observe that 
the size of the denser cloud is independent of large N, which implies that the 
atoms are being self-trapped in the non-local minimum. 

For the case of a spherical trap with trap size $a_{0}=3.0 \mu m$, we fail 
to find the metastable numerical solutions beyond N $\simeq1430$ atoms. The 
instability occurs as the number of atoms increases, the net attractive 
interaction between atoms becomes dominant and kinetic energy can no longer 
stabilize the wave function. In the mean-field approach it has already been 
pointed out that the transition to an unstable state may occur due to quantum 
tunnelling. This tunnelling is similar to the ordinary quantum tunnelling 
and its rate can be calculated by the semiclassical formula. As this rate 
is quite small, neglecting tunnelling we consider only the loss rates due 
to two-body dipolar collisions and three-body recombination collisions. 

\vskip 0.5cm
\begin{figure}[hbpt]
\vspace{-10pt}
\centerline{
\hspace{-3.3mm}
\rotatebox{0}{\epsfxsize=9cm\epsfbox{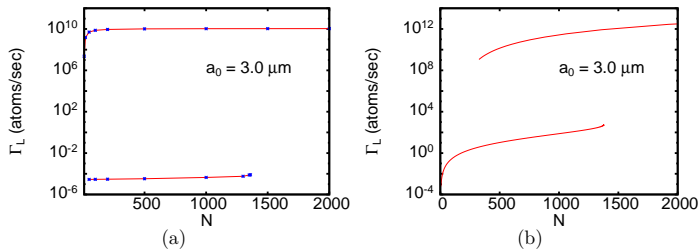}}}
\caption{(Color online) Loss rate due to two- and 
three-body collisions as a function of 
the number of $^{7}$Li atoms using PHEM [panel (a)] and GP with non-local 
interaction [panel (b)].}
\end{figure}

The total loss rate is given by
\begin{equation}
 \Gamma_{L}=K \int d^{3}\vec{x}|\phi(\vec{x})|^{4} + L \int d^{3}\vec{x}
|\phi(\vec{x})|^{6}
\end{equation}
where two-body dipolar loss rate coefficient $K=1.2\times10^{-2}\mu 
{\rm m}^{3}{\rm sec}^{-1}$ and the three-body recombination loss rate 
coefficient $L=2.6\times10^{-4}\mu {\rm m}^{6}{\rm sec}^{-1}$~\cite{moerdijk}. 
$\phi(\vec{x})$ is the condensate wave function. 
This is simply given by Eq.~(20) for the non-local GP approach. For the 
many-body approach, the wave function $\psi$ is initially obtained as 
a function of the Jacobi coordinates. It is then transformed 
into a function 
of position vectors $\{{\vec x}_1,{\vec x}_2, ..., {\vec x}_A\}$ of 
$A$ particles. Finally, the one-body density is obtained by 
integrating $|\psi|^2$ over all ${\vec x}_i$, except one. This 
one-body density is used for $|\phi({\vec x})|^2$ in Eq.~(25). 
It is already known that in case of a negative scattering length, the rapid 
increase in the density of atoms and large increase in the net attractive 
interaction for a large number of atoms, result in a very large loss rate 
near the critical point. Thus for the unique metastable branch in GP 
mean-field picture with local interaction, the loss rate increases very 
fast with the increase in the number of 
atoms [7]. However, due to the presence of the deep attractive well in the 
many-body picture, we calculate the loss rate for both the branches and 
observe the discontinuity. By using condensate wave function as described 
above, we calculate $\Gamma_{L}$ 
by using Eq.~(25) and the results are plotted in panel (a) of Fig.~4. 
The lower branch corresponds to the metastable region. As the metastable 
region is flatter compared to the NAW, although the atoms are correlated, 
the probability of two or more atoms to come very close is relatively small. 
Thus the loss rate is initially small and increases with increase in atom 
number. The upper branch accounts for the loss rate in the deep narrow 
attractive well. As all the atoms are now confined within a very narrow 
and deep well, they suffer vigorous two- and three-body collisions and 
the calculated loss rate is much higher than the same calculated in the 
metastable region. The transition between these two branches is again 
discontinuous. In panel (b) of Fig.~4 we plot $\Gamma_{L}$ calculated 
using variational trial wave function of Eq.~(20) using non-local interaction. 
Substitution of trial wave function Eq.~(20) in Eq.~(25), leads to 
\begin{equation}
 \Gamma_{L}(N)=\frac{KN^{2}}{(2\pi)^{3/2}a_{0}^{3}\sigma^{3}} + 
\frac{LN^{3}}{(3\pi^{2})^{3/2}a_{0}^{6}\sigma^{6}}.
\end{equation}
In this formula, N is related to $\sigma$ through Eq.~(22). The existence 
of low-density metastable branch and high-density stable branch which are 
separated by an intermediate unstable branch is also reflected in Fig.~4. 
It is in sharp contrast to the existence of a unique metastable branch in 
the mean-field approach with local potential. For the sake of completeness, 
in Fig.~5 we plot the loss rate for traps of different sizes.

\vskip 0.5cm
\begin{figure}[hbpt]
\vspace{-10pt}
\centerline{
\hspace{-3.3mm}
\rotatebox{0}{\epsfxsize=9cm\epsfbox{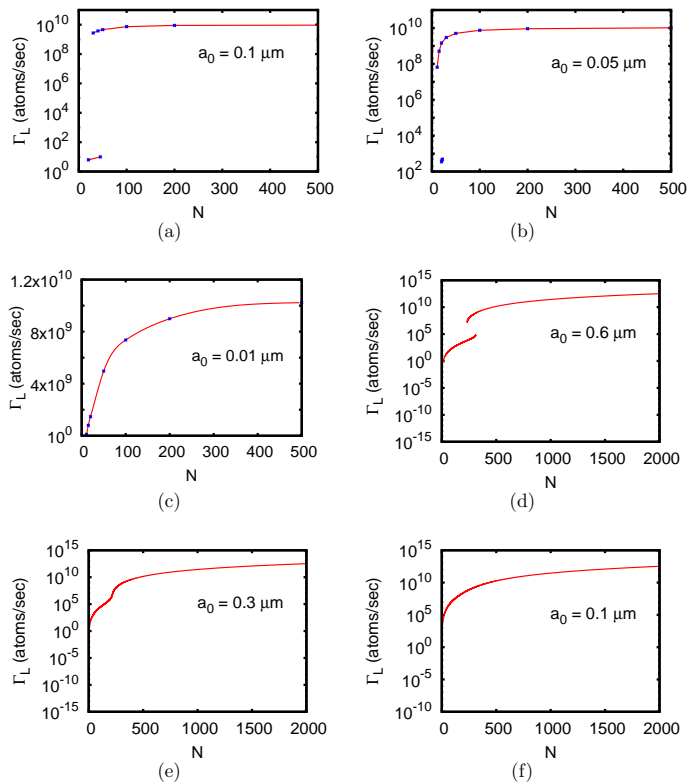}}}
\caption{(Color online) Loss rate due to two- and three-body collisions 
as a function of the number of $^{7}$Li atoms using 
PHEM [panels (a) -- (c)] and GP with 
non-local interaction [panels (d) -- (f)]. Note that in panels (a) - (c), the horizontal
axis is upto $N=500$, to show the small $N$ part prominently. Calculated
values of $\Gamma_L$ for $N$ above $500$ remain practically constant for PHEM.}
\end{figure}

\section{Conclusions}

In summary, we have analyzed the stability of the attractive 
Bose-Einstein condensate 
of $^7$Li atoms in a harmonic trap using an \textit{ab initio} quantum 
many-body calculation incorporating realistic van der Waals interaction 
and all possible two-body correlations. Unlike the single metastable 
condensate arising from the GP mean-field theory with local interaction, 
we observe that the atomic cloud may exist in different states. 
In addition to the commonly known low density metastable BEC state, we 
find a high density stable solution, which corresponds to the 
cluster state of the lithium atoms. Between these two minima, there is 
a maximum of the effective potential, which does not 
correspond to any stable physical state of the Bose gas. The local 
GP approach produces only the metastable solution, together with an 
intermediate barrier (local maximum), but the effective potential 
has no lower bound on the left of the intermediate barrier. Indeed 
there is a pathological singularity at the origin, which only 
produces a pathological collapse (not leading to any physical solution). 
The effect of a non-local momentum dependent interaction (replacing 
the local contact interaction) in the GP approach was investigated 
in earlier works~\cite{luca} using a variational {\it ansatz}. The 
results of this non local GP approach agree qualitatively with our 
present observations. 
The lack of quantitative agreement is attributed to the difference 
in the origins of non-locality in the two approaches. 
We have investigated the average size of the 
bosonic system, as also the loss rates due to two- and three-body 
collisions for both the metastable BEC state and the collapsed 
cluster state. In general these quantities lie on separate 
branches for the two states. For larger trap lengths, the two branches 
are unconnected. But there is a continuous transition for very tight 
traps. The local GP approach produces only one branch, corresponding 
to the metastable state. The local GP uses an attractive contact 
interaction for the $^7$Li condensate. This causes a pathological 
singularity at the origin, preventing any stable collapsed state. 
The non local GP approach uses a non-local potential, which is 
attractive at larger separations, but has a repulsive contact 
interaction as $r \rightarrow 0$. Hence there is no pathological 
singularity and the effective potential becomes strongly repulsive 
as $r \rightarrow 0$. This gives rise to the stable high density 
branch. In the PHEM approach, the realistic van der Waals (vdW) 
potential is used. This has a very strong repulsion at short 
separations and an attractive $1/|{\vec x}_{ij}|^6$ long 
tail. The strongly 
repulsive core of the vdW potential, together with the 
hypercentrigugal repulsion, arising from the PHEM approach, 
produces a repulsive core followed by a deep attractive well 
as one moves away from the origin in the effective potential, 
$\omega_0(r)$. Thus both the PHEM and the non local GP approaches 
produce the same qualitative results. However, 
quantitative differences 
exist due to different origins of non-locality in these two approaches. 
While, non-locality in the many-body approach arises from the use of 
vdW potential with a repulsive core and the hypercentrifugal repulsion, 
that in the non local GP approach is built in the choice of the 
two-body potential. 

Although the presence of an intermediate state has already been established in 
earlier works where GP mean-field theory has been employed with 
non-local interaction~\cite{luca}, the effect of the short-range behavior of 
realistic interatomic potential on the intermediate state has not 
been investigated as yet. Especially in the collapse regime, where 
the atomic cloud becomes highly correlated, the use of a correlated 
many-body approach is required for the accurate determination of 
the collapse point and various properties of the atomic cluster. 
The non-local interaction in the effective interaction can prevent 
a pathological 
collapse, which results in the GP approach with a purely local attractive 
interaction. The comparison of these two 
approaches is elaborately discussed in the present work.
In the usual trap setup we observe a discontinuous jump 
between the low-density and high-density phases. However, the use of 
tight traps is also interesting and we observe how the discontinuous 
jump can be avoided by using microtraps. 
For large number of atoms, 
the system remains in a quantum self-trapped state.

\begin{center}
{\bf Acknowledgements}
\end{center}

This work is supported in part by DST (Grant No. SR/S2/CMP-59-2007); 
CSIR (Grant No. 09/028(0773)-2010-EMR-1); UGC (Grant No. F.6-51/(SC)/2009 
(SA -II)); DAE (2009/37/23/BRNS/1903).

\end{document}